# A simulation framework for statistical inference on the alerting capabilities of smartphone-based earthquake early warning systems. With a case study on the Earthquake Network system in Haiti

Francesco Finazzi[1,*], Frank Yannick Massoda Tchoussi[1]

[1]University of Bergamo, Department of Economics; Bergamo, Italy.
*Corresponding author. Email: francesco.finazzi@unibg.it

**ABSTRACT**: smartphone-based earthquake early warning systems implemented by citizen science initiatives are characterized by a significant variability in their smartphone network geometry. This has an direct impact on the earthquake detection capability and performance of the system. Here, a simulation framework based on the Monte Carlo method is implemented for making inference on relevant quantities of the earthquake detection such as the detection distance from the epicentre, the detection delay and the warning time for people exposed to high ground shaking levels. The framework is applied to Haiti, which has experienced deadly earthquakes in the past decades, and to the network of the Earthquake Network citizen science initiative, which is popular in the country. It is discovered that relatively low penetrations of the initiative among the population allow to offer a robust early warning service, with warning times up to 12 second for people exposed to intensities between 7.5 and 8.5 of the modified Mercalli scale.

## 1. Introduction

Since 2013, the Earthquake Network (EQN) citizen science initiative implements the first earthquake early warning (EEW) system based on smartphone networks (Finazzi, 2016). Similarly to other initiatives (Kong et al. 2016), accelerometers on-board smartphones are exploited to measure the ground shaking, for detecting earthquakes in real-time and sending alerts to people not yet reached by the damaging seismic wave. The earthquake detection logic is implemented on a central server which collects the information coming from the smartphones and which, thanks to a statistical algorithm (Finazzi and Fassò, 2017), is able to understand if an earthquake is occurring. Bossu et al. (2021) proved that EQN has similar detection capability of the more expensive EEW systems based on scientific-grade instruments (Kohler et al. 2018), while Fallou et al. (2021) showed that the EQN alerting service is highly appreciated by citizens.

EQN's smartphone network geometry is highly dynamic. People opt-in and opt-out from the initiative at any time, plus the number of monitoring smartphones exhibits an intra-day variability due to people charging their smartphones mostly at night. In Finazzi et al. (2022), it is shown that the number of monitoring smartphones affects the earthquake probability of detection and the detection delay, which, in its turn, affects the warning time for people exposed to a given shaking level.

In this work, a simulation framework based on the Monte Carlo method is deployed to study the impact of the network geometry on the expected warning time for citizens who experience a life-threatening earthquake. Inputs of the simulation framework are the spatial distribution of the monitoring smartphones, the population spatial distribution and the spatial distribution of the earthquake intensity. The main output consists of the warning time distributions on people exposed to some ranges of ground shaking intensity. To facilitate statistical inference, all outputs are provided with measures of uncertainty.

The simulation study is applied to Haiti which, in its recent history, has witnessed two catastrophic earthquakes. The first on January 12, 2010 with more than 100,000 casualties and the second on August 14, 2021 with 2248 casualties.

## 2. Simulation framework

The simulation framework is based on the Monte Carlo method (Rubinstein and Kroese, 2016). Reason for this choice is that the links between smartphone network geometry, detection delay, detection location and warning time are non-trivial and hardly formalizable in equations.

Considering a major earthquake, the framework aims at:

1. Estimating the EQN detection delay distribution (with respect to origin time).
2. Estimating the EQN detection separation distribution (with respect to the epicentre).
3. Estimating the warning time distribution for people exposed to high shaking levels.

Simulation inputs taken into account by the framework are:

1. The population spatial distribution.
2. The spatial distribution of the modified Mercalli intensity (MMI).
3. The smartphone network geometry.

Inputs 1. and 2. are assumed to be deterministic while 3. is stochastic. Indeed, the network may be described as a stochastic object since smartphones become active at random locations and random times, and the length of time a smartphone stays active is also random.

Under the Monte Carlo method, the EQN earthquake detection is repeatedly simulated using the detection algorithm detailed in Finazzi and Fassò (2017), conditionally on a network geometry which changes at each Monte Carlo run. This induces distributions on detection delay, detection separation and warning time.

## 3. Case study

The simulation framework outlined in the previous section is applied to Haiti where EQN works as a public EEW system with around 10,000 daily active users. In the next sections, each simulation input is detailed, as well as the EQN detection algorithm.

### 3.1. Population spatial distribution

Population distribution in Haiti is needed to assess the population exposure to any given value of MMI and to assess the (statistical) distribution of the warning time for people exposed to such MMI value.

For this work, we adopted the Gridded Population of the World collection (CIESIN, 2018) in its 4th version, and in particular the Population Count product for year 2020 at 30 arc-second (around 1 km) spatial resolution. The 2020 population distribution is used as a proxy of today's population.

### 3.2. Earthquakes and MMI spatial distribution

Major earthquakes are characterized by finite fault ruptures (Böse et al. 2012, Goda 2019) and complex focal mechanisms (Kagan 2017) which lead to spatially anisotropic intensity fields. Intensity fields are hard to predict and they are usually assessed a-posteriori.

In order to provide realistic EQN performance during a major earthquake, we base our analysis on two deadly past events with epicentres in Haiti and for which intensity fields are available.

The first event occurred on January 12, 2010 with magnitude 7.0 while the second event occurred on August 14, 2021 with magnitude 7.2. Figure 1 shows the spatial distribution of the MMI (USGS 2010, 2021) for the two events while Figure 2 depicts the MMI population exposure assuming the population distribution of 2020. Despite the lower magnitude, the 2010 event had an epicentre much closer to the metropolitan area of Port-au-Prince and this is why, if the same event happened today, the number of people exposed to high values of MMI would be high.

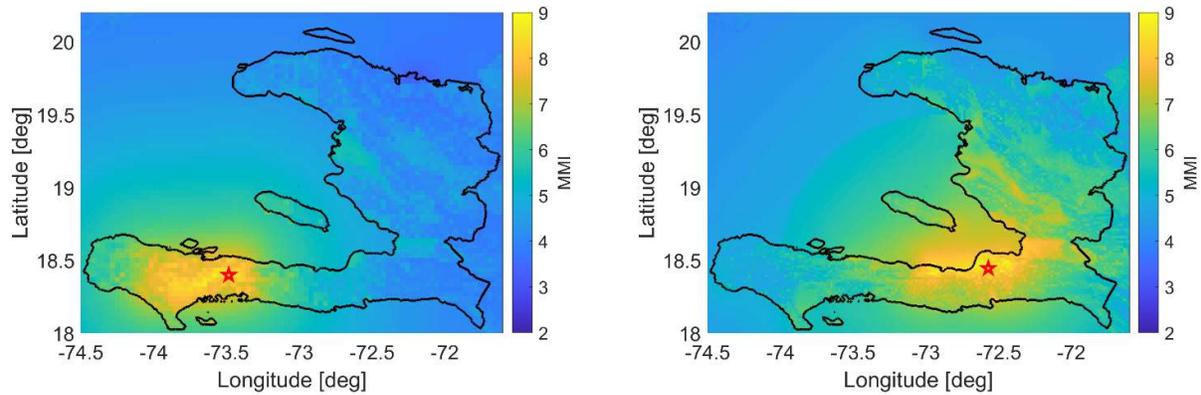

*Figure 1. Modified Mercalli Intensity maps for the M7.2 event on August 14, 2021 (left) and for the M7.0 event on January 12, 2010 (right). Source: USGS.*

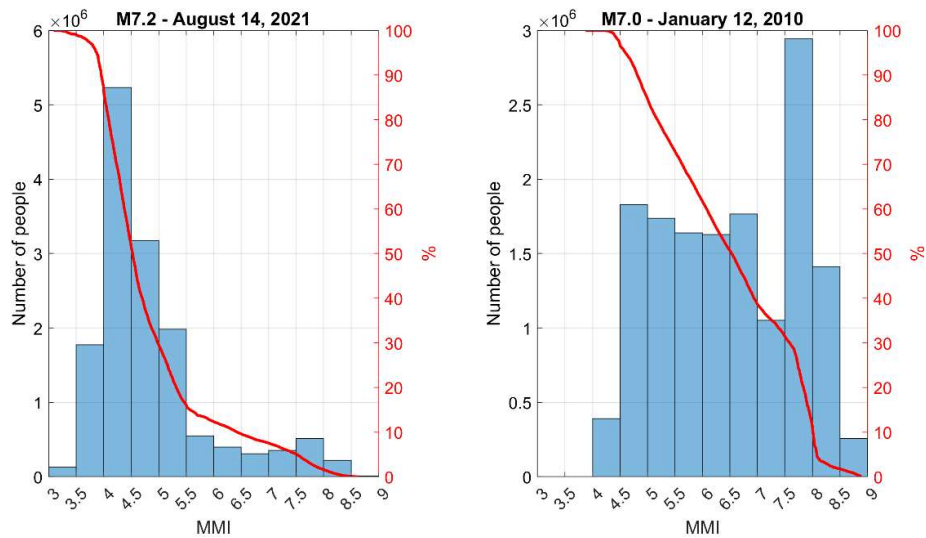

*Figure 2. Histograms of the MMI population exposure for the M7.2 event on August 14, 2021 (left) and for the M7.0 event on January 12, 2010 (right) assuming the population distribution in 2020. Red curves give, for each MMI value, the percentage of the population that was exposed to an intensity equal or higher than the MMI value.*

### 3.3. Network simulation

Simulating the smartphone network at the time of the earthquake is a crucial aspect of this work since the network geometry directly affects the EQN detection capabilities and performance.

To simulate realistic network geometries, we rely on the spatial coordinates of N=6202 EQN users living in Haiti and with a functioning smartphone app (namely an app that has been monitoring for earthquake at least once since it was installed).

The N coordinates are potential smartphone locations that we use to simulate the network geometry. At any given time, however, the number of smartphones monitoring is much smaller than N since the EQN app only works when the smartphone is not moving and it is charging. Figure 3 shows the trend of the number of monitoring smartphones in Haiti during a given day. With respect to the daily average, the intra-day variation is around 50% in positive and in negative.

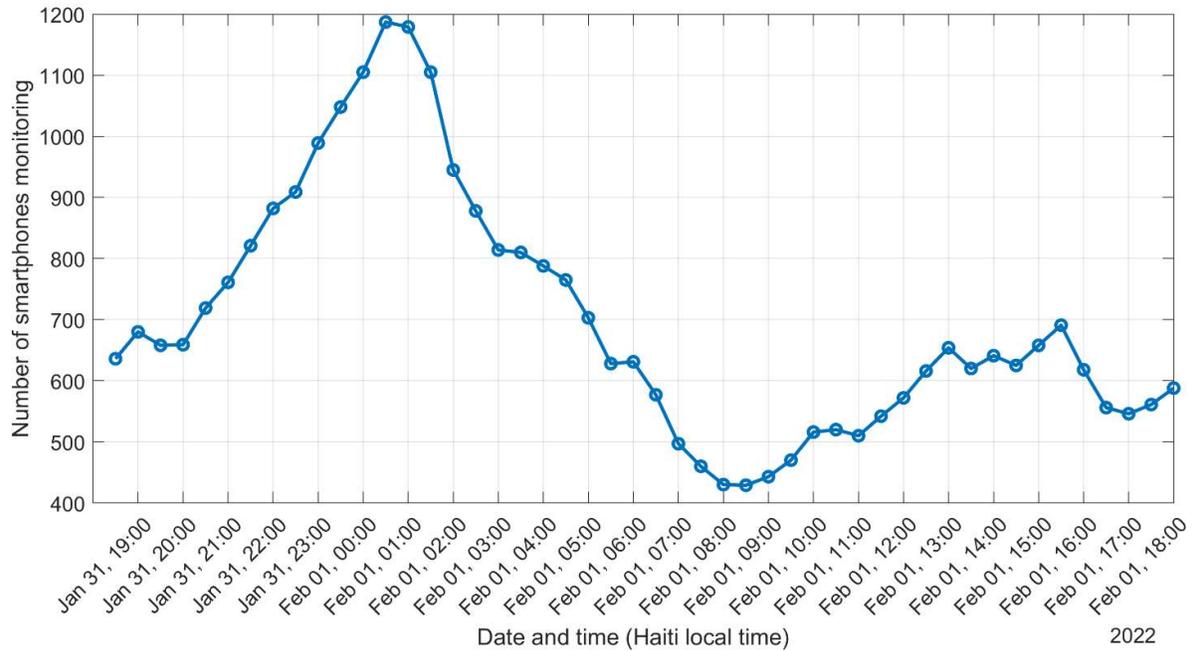

*Figure 3. Daily trend of the number of EQN smartphones monitoring for earthquakes in Haiti from January 31, 18:30 to February 1, 18:00 Haiti local time.*

For this reason, we condition the simulation of the network geometry to the number of active smartphones, that we fix to one of these possible values: $n$ = (300, 400, 500, 600, 700, 800, 900, 1000, 1100, 1200, 1300, 1400, 1500, 1600, 1700, 1800, 1900, 2000, 2100, 2200, 2300, 2400, 2500, 2600, 2700, 2800, 2900, 3000).

For a given $n$, the network geometry is simulated by randomly sampling (without replacement) n out of the N spatial coordinates. When $n$ is small, the network is sparse and highly variable across different simulations while, when n is large, the network is dense and slightly variable. Figure 4 shows two simulated network geometries when $n$ is 200 and 3000.

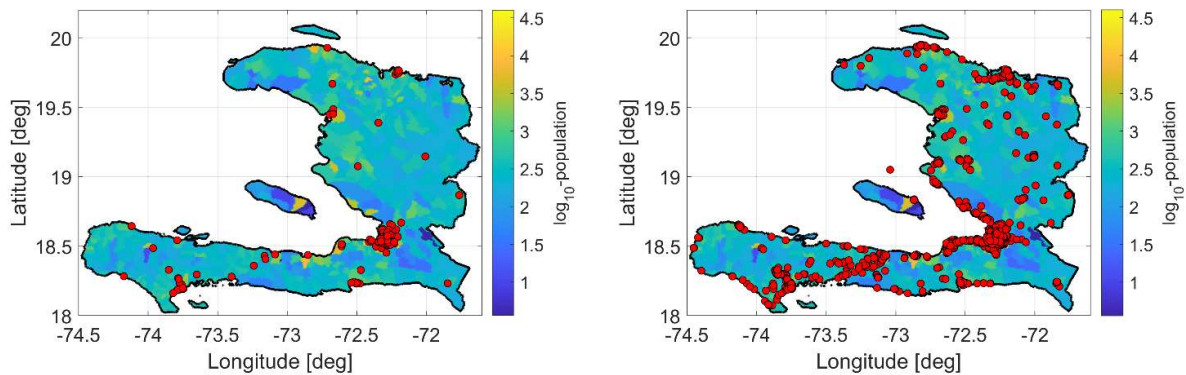

*Figure 4. Simulated EQN network geometry with 300 smartphones (left panel) and with 3000 smartphones (right panel). Each dot represents the simulated location of a monitoring smartphone.*

### 3.4. Earthquake detection simulation

The detection of an earthquake by EQN is simulated in two steps. At the first step, conditionally on the simulated smartphone network (see the previous section), we simulate the detection of the earthquake by each single smartphone. The following assumptions are made:

1. Smartphones detect the P wave.
2. Smartphones have a random detection delay described by a uniform distribution with support [0.5, 3.5] s.
3. Smartphones have a 0.7 probability to detect the earthquake and send the information to the EQN server.

Assumption 1. comes from Bossu et al. (2021) which shows that EQN detects the P wave when the earthquake magnitude is above 5. Assumption 2. depends on the detection algorithm implemented on the smartphone and on the Internet latency while assumption 3. acknowledges that smartphones may be unable to detect the earthquake or unable to send the information to the server when the earthquake strikes.

The output of the first step is the list of smartphones that detected the earthquake. Each list entry includes the smartphone coordinates (latitude and longitude) and the smartphone detection time.

At the second step, the server-side EQN detection algorithm (Finazzi and Fassò, 2017) is applied to the output of the first step. The output of the second step is the detection location (which is a preliminary estimate of the earthquake epicentre) and the detection time.

Detection location and detection time directly depends on the network geometry at the time of the earthquake. By simulating a large number of random network geometries, we are able to study how location and time change, to estimate the expected location and time, and to provide confidence intervals on both estimates.

## 4. Simulation results

Simulation results presented in this section are based on a Monte Carlo simulation with 1000 replica (of random network geometry) for each network size *n* and each of the two earthquake events.

The first simulation result discussed here is the EQN detection delay and detection distance with respect to the number of monitoring smartphones (Figure 5 and Figure 6). In general, the higher the number of smartphones the lower the detection delay and detection distance. Nonetheless, after around 1200 monitoring smartphones the gain is little. Note that 1200 monitoring smartphones is three times the minimum value of the graph of Figure 3, meaning that, on February 1, 2022, the EQN app penetration in Haiti was three times lower than an ideal penetration that would have guaranteed a nearly stable EQN performance over the 24 hours of the day.

Figure 7 and Figure 8 show the Monte Carlo results in terms of spatial variability of the detection location assuming the epicentres of the August 14, 2021 and of the January 12, 2010 events, respectively. Conditionally to a network geometry with 300 and with 3000 monitoring smartphones, the 1000 detection locations obtained from the Monte Carlo replicas were used to estimate the spatial density function that describes the probability that the EQN detection occurs in a given area. The density function is more compact in space when the number of smartphones is high (and the variability of the network geometry is low). This means that, when the number of smartphones is high, the behaviour of the EQN network is more predictable.

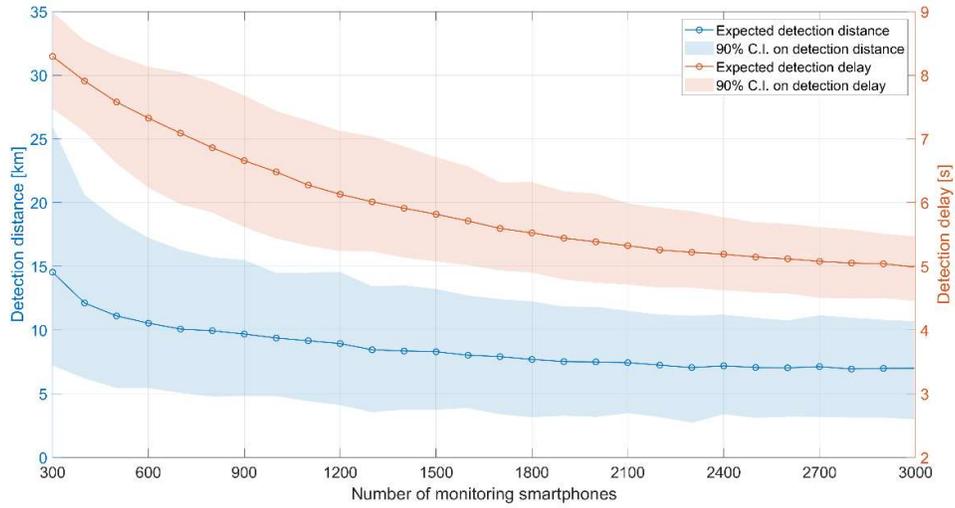

*Figure 5. EQN detection distance and detection delay vs number of monitoring smartphones assuming the earthquake epicentre of the August 14, 2021 event. Solid lines are the Monte Carlo averages while the coloured areas depict the Monte Carlo 95% confidence bands.*

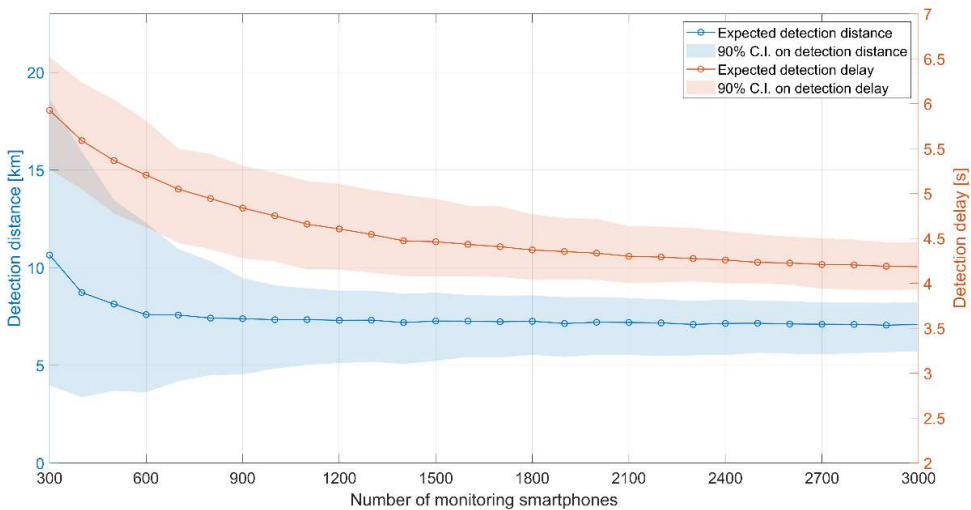

*Figure 6. EQN detection distance and detection delay vs number of monitoring smartphones assuming the earthquake epicentre of the January 12, 2010 event. Solid lines are the Monte Carlo averages while the coloured areas depict the Monte Carlo 95% confidence bands.*

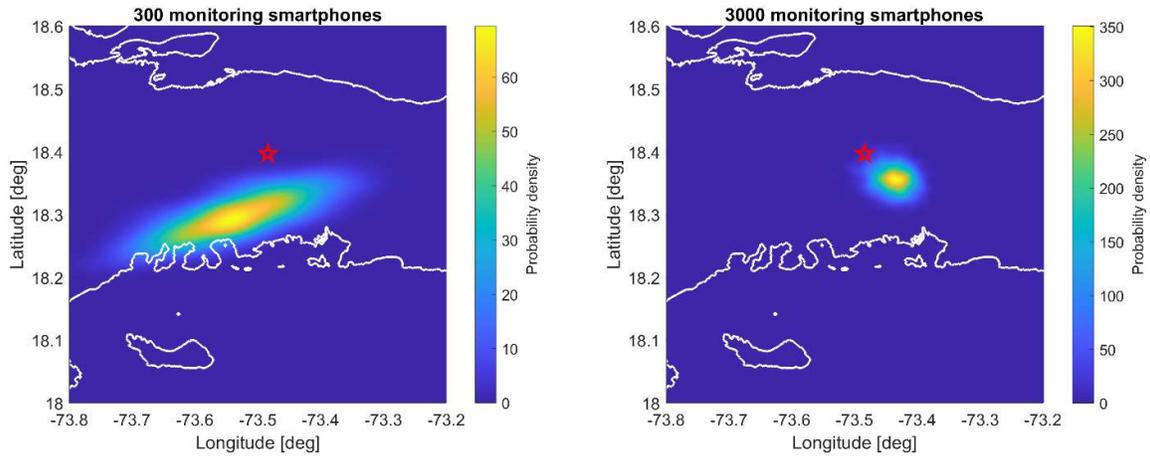

*Figure 7. Spatial density functions of the EQN detection location assuming a network geometry with 300 monitoring smartphones (left panel) and with 3000 monitoring smartphones (right panel) assuming the epicentre (red star) of the August 14, 2021 event.*

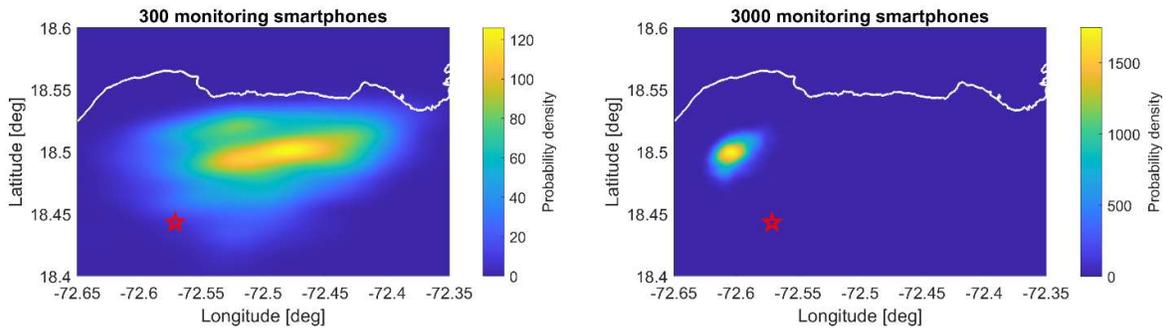

*Figure 8. Spatial density functions of the EQN detection location assuming a network geometry with 300 monitoring smartphones (left panel) and with 3000 monitoring smartphones (right panel) assuming the epicentre (red star) of the January 12, 2010 event.*

Figure 9 and Figure 10 show, for different ranges of MMI, the histograms of the hypothetical warning time distribution among the Haitian population if the two earthquake events are detected by EQN at the expected detection locations (namely the modes of the spatial density functions shown in Figure 7 and Figure 8). The word "hypothetical" is used to stress that instantly delivering the alert to millions of people is not a trivial problem and that the actual warning time may be lower.

Note that, in many cases, the warning time in the histograms of Figure 9 and Figure 10 is largely variable. This is mainly due to the fact that the MMI spatial distribution is highly anisotropic. People exposed to the same MMI are not necessarily at the same distance from the epicentre. In particular, for a given MMI, people living in the same direction of the fault rupture benefit from a larger warning time.

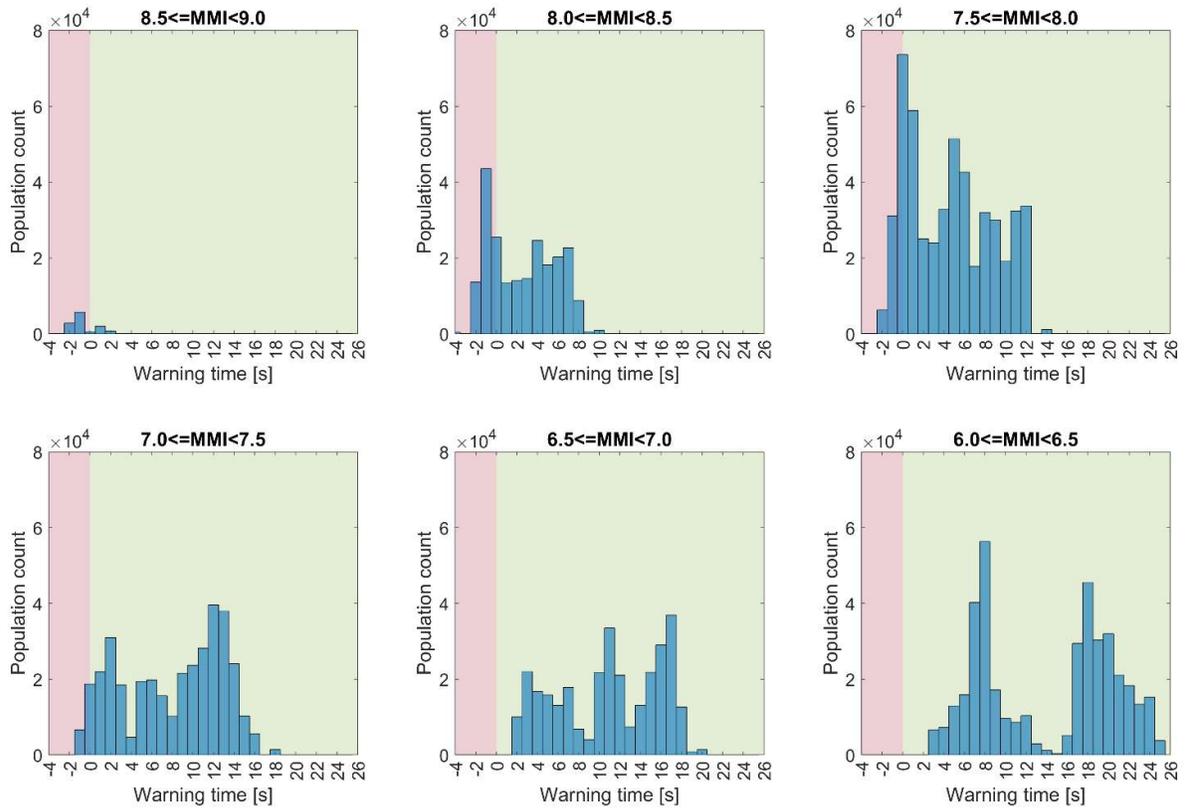

*Figure 9. Warning time distribution over the Haitian population assuming the MMI spatial distribution of the August 14, 2021 event and an EQN detection at the expected detection location when the number of monitoring smartphones is 3000.*

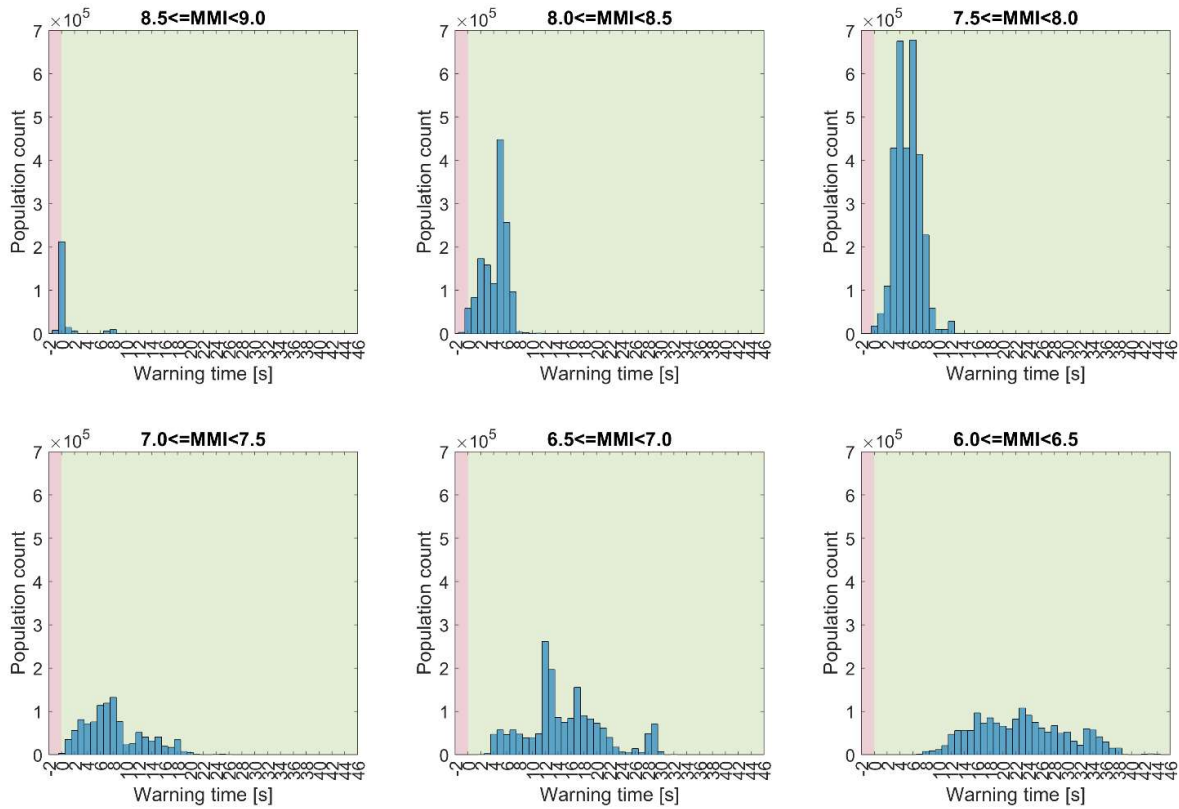

*Figure 10. Warning time distribution over the Haitian population assuming the MMI spatial distribution of the January 12, 2010 event and an EQN detection at the expected detection location when the number of monitoring smartphones is 3000.*

The warning time distributions in Figure 9 and Figure 10 refer to an EQN detection at the expected detection location and to a smartphone network with 3000 monitoring smartphones. We are interested in understanding how those distribution change when the number of smartphones and the detection location also change.

To synthetize the warning time distributions in an easy and interpretable manner, we consider for each distribution (histogram) the 2.5[th] percentile, the average and the 97.5[th] percentile of the distribution. For each MMI range, the average represents the average warning time for people exposed to a MMI in that range. The 2.5[th] percentile represents the warning time for the "unfortunate" people who are exposed to a high MMI and a low (or even negative) warning time, while the 97.5[th] percentile represents the warning time for the "lucky" people who are exposed to a high MMI and a high warning time. Figure 11 and Figure 12 show, for some MMI ranges, how these quantities evolve with respect to the number of monitoring smartphones and what is their variability due to variations in the detection location.

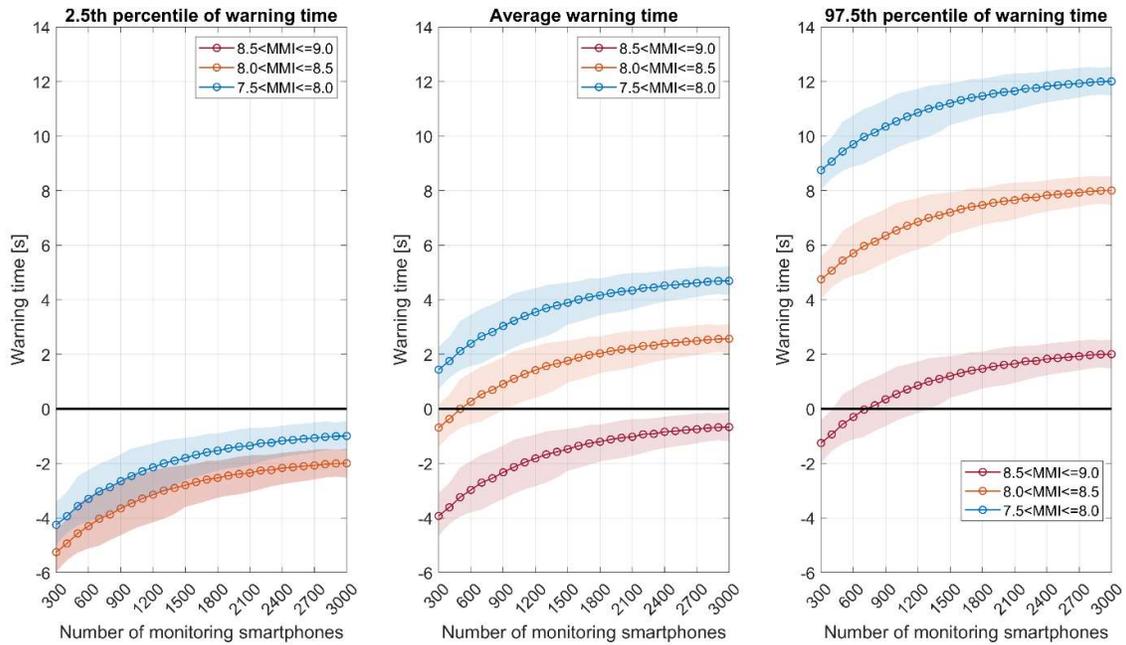

*Figure 11. 2.5th percentile (left panel), average (centre panel) and 97.5th percentile (right panel) of the warning time for people exposed to a MMI in the range (8.5, 9], (8, 8.5] and (7.5, 8] for different numbers of monitoring smartphones assuming the MMI spatial distribution of the August 14, 2021 event. Solid lines are the Monte Carlo averages while the shaded areas depict the 95% confidence bands.*

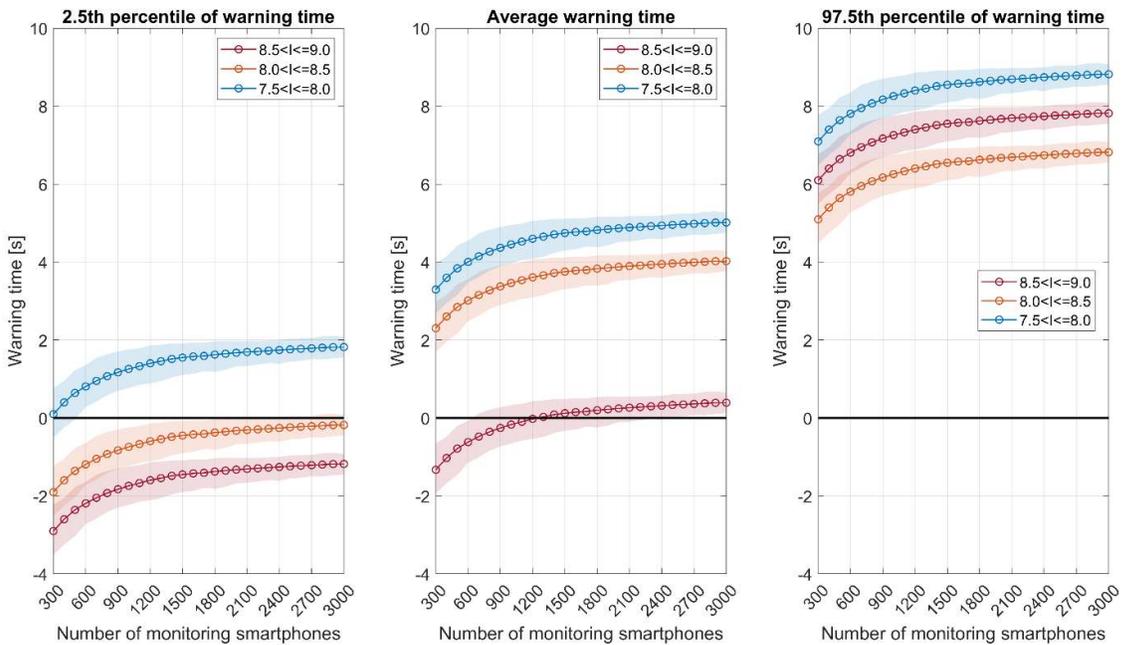

*Figure 12. 2.5th percentile (left panel), average (centre panel) and 97.5th percentile (right panel) of the warning time for people exposed to a MMI in the range (8.5, 9], (8, 8.5] and (7.5, 8] for different numbers of monitoring smartphones assuming the MMI spatial distribution of the January 12, 2010 event. Solid lines are the Monte Carlo averages while the shaded areas depict the 95% confidence bands.*

## Discussion and conclusions

Smartphone-based EEW systems are characterized by a network which is dynamic in the number of smartphones and in their spatial distribution. This affects the system detection capability and its usefulness for people in terms of warning time. In a context of inferential statistics, this work developed a simulation framework for assessing the impact of network geometry on the warning time distribution for people exposed to high levels of ground shaking.

The framework was applied to Haiti using as reference two past deadly earthquakes for which the intensity spatial distribution is known. We discovered that:

1. If similar (in magnitude and epicentre) events occurred today, EQN would be able to issue alerts with a positive warning time (up to 12 seconds) for people exposed to dangerous shaking levels (e.g., above intensity 7.5).
2. EQN performance tends to stabilize when the number of monitoring smartphones (in Haiti) is above around 1200. This corresponds to around 1/10,000 of the Haitian population.
3. Within any given intensity class (e.g., 7.5≤MMI<8.0), the distribution of the warning time is highly variable. This is a consequence of the anisotropy of the intensity spatial distribution coupled with the population distribution.
4. For any fixed number of monitoring smartphones, the uncertainty on their spatial distribution does not largely affect the uncertainty on the warning time (see confidence intervals in Figure 11 and Figure 12).

These results suggest that, in Haiti, robust low-cost smartphone-based EEW is possible starting from low penetrations of the EQN smartphone app on the population. This could benefit the many ongoing citizen science initiatives in Haiti (Calais et al. 2022; Fallou et al. 2022).

## Funding

This article was funded by the European Union's Horizon 2020 Research and Innovation Program under grant agreement RISE No. 821115. Opinions expressed in this article solely reflect the authors' views; the EU is not responsible for any use that may be made of information it contains.

## Competing Interests

The authors have no relevant financial or non-financial interests to disclose.